\documentclass{svjour2}                    % onecolumn
\smartqed  % flush right qed marks, e.g. at end of proof
\usepackage{textcomp}
\usepackage{graphicx}
 \journalname{J. Low. Temp. Phys.}
\begin{document}

\title{Effects of charging energy on SINIS tunnel junction thermometry%\thanks{Grants or other notes
}


\author{P.~J. Koppinen         \and
        T. K\"uhn							\and
        I.~J. Maasilta %etc.
}


\institute{%F. Author \at
              Nanoscience Center, Department of Physics, P.O.Box 35, FI--40014 University of Jyv\"askyl\"a, Finland\\
              Tel.: +358-14-260 4720\\
              Fax: +358-14-260 4756\\
              \email{panu.koppinen@jyu.fi}           %  \\
}

\date{Received: date / Accepted: date}

\maketitle

\begin{abstract}
We have investigated theoretically the effects of the charging energy to the normal metal--insulator--superconductor (NIS) tunnel junction used as a thermometer. We demonstrate by numerical calculations how the charging effects modify NIS thermometry, and how the voltage--to--temperature response  and the responsivity $|\mathrm{d}V/\mathrm{d}T|$ of a current biased thermometer are affected. In addition, we show that the responsivity of the thermometer can be modulated with an additional gate electrode. The maximum responsivity is achieved when the Coulomb blockade is maximal, i.e. with a closed gate.
\end{abstract}

\section{Introduction}
\label{intro}
Tunnel junction thermometry with normal metal--insulator--superconductor (NIS) junctions was discussed a while 
back\cite{rowellthermometry}, but has only recently been widely used in low--temperature thermal transport experiments 
\cite{Leivomembrane,Clelandtransport,Photon,karvonenPRL}. Advantages of NIS thermometry are e.g. low self--heating, good 
responsivity, ease of integration into the system under study, small size, existing high--frequency read--out 
schemes\cite{RFSINIS} and the fact that it can be used as local probe for temperature in nanostructures.
Ideally, NIS junctions can be considered as primary thermometers, since their voltage--to--temperature (or 
current--to--temperature) calibration curves are determined by only two parameters: the superconducting gap $\Delta$ and the 
tunneling resistance $R_{T}$, which can be determined quite accurately from the measured current--voltage ($I$--$V$) 
characteristics of the junction. However, charging effects\cite{singlecharge} complicate this picture for small area (capacitance) tunnel junctions connected to small islands (small self--capacitance). In this case charging effects influence the voltage(or current)--to-temperature response of the thermometers, if the charging energy $E_{C}=e^2/(2C)$ is of the same order of magnitude or larger than temperature, i.e. $E_{C}\geq T$. This limit is easily obtainable in small devices, such as the recently demonstrated heat transistor\cite{heattransistor} or the hybrid single electron transistor\cite{SINISSET1,SINISSET2}, where charging energies  $E_{C} \sim $ 1 K have been demonstrated. In this paper we investigate theoretically how $E_{C}$ affects the current biased voltage--to--temperature responsivity of the SINIS thermometer and how it can be tuned with an additional gate located in the close proximity of the junctions. For simplicity, we only concentrate on the case, where single particle tunneling is taken into account, and higher order two--particle (Andreev) processes\cite{Hekking,courtois} are left out in the discussion.
In addition to the charging effects, we also discuss the effect of the non--ideal single particle current caused by the broadened %%@
density of states due to the finite life--time of quasiparticles in the superconductor\cite{Dynes}, which causes deviations from %%@
the ideal behavior in thermometry.

\section{SINIS thermometry without charging effects}
\begin{figure*}[ht]
\includegraphics[width=\textwidth]{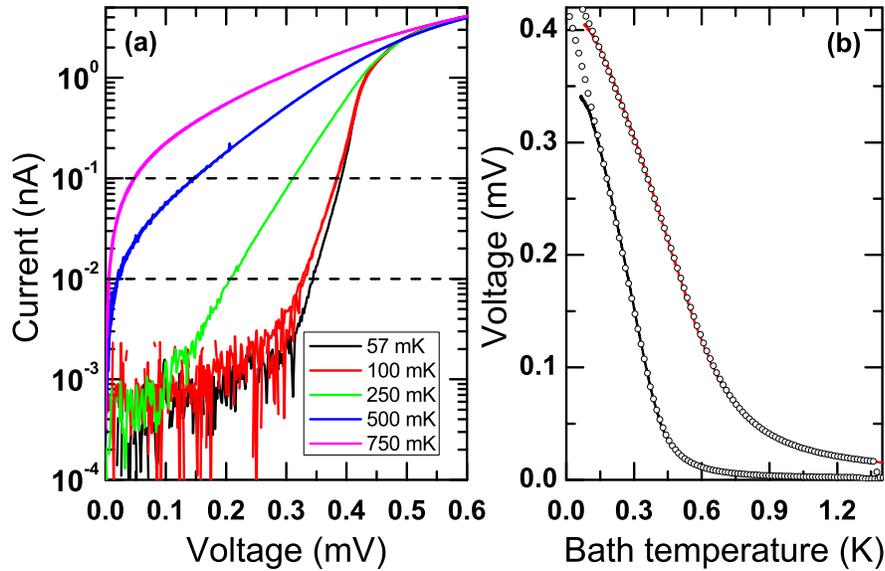}
\caption{\label{ExptIV}(Color online)(a) Typical measured sub--gap current--voltage characteristics of a SINIS probe junction at %%@
different temperatures (from right to left, lowest temperature on the right). Dashed horizontal lines from top to bottom correspond %%@
to bias currents of 100 pA and 10 pA, respectively. 
(b) SINIS thermometer voltage vs. bath temperature, black (bottom curve) and red (top curve) lines corrrespond to two different %%@
current bias points 10pA and 100 pA, respectively. Open circles represent the calculation from Eq. %%@
(\ref{current_sym}).}
%\begin{minipage}[t]{14pc}
%\includegraphics[height=7cm]{SF1}
%\caption{\label{SF1}(Color online)Typical measured sub--gap current--voltage characteristics of SINIS probe junction at %%@
%different temperatures (from left to right, lowest temperature on left). Solid horizontal lines from top to down correspond %%@
%currents 100 pA and 10 pA, respectively.}
%\end{minipage}\hspace{5pc}
%\end{figure}
%\begin{minipage}[t]{14pc}
%
%\includegraphics[height=7cm]{calibration}
%\caption{\label{SF2}(Color online) Calibration curves of SINIS thermometer, black (bottom curve) and red (top curve) lines %%@
%corrrespond two different current bias points 10pA and 100 pA, respectively. Open circles represent the BCS--theory %%@
%calculation.}
%\end{minipage} 
\end{figure*}

The non--linear current--voltage ($I$-$V$) characteristics (Fig. \ref{ExptIV} (a)) of a NIS tunnel junction can be used for %%@
thermometry (see e.g. \cite{Jukkareview} for a good review). In practice, thermometry is typically carried out by current biasing %%@
the junction at a constant current and measuring the voltage response of the thermometer, which is only a function of temperature. %%@
To obtain a larger signal, i.e. increased responsivity for thermometer, a structure containing two junctions in series (SINIS) can %%@
be used instead of a single NIS junction. Measured $I$--$V$ curves for a typical SINIS thermometer with $E_{C}<<k_{B}T$ at %%@
different bath temperatures are shown in Fig. \ref{ExptIV} (a).  It can be seen from Figs. \ref{ExptIV} (a) and (b) that the %%@
voltage responsivity $\mathrm{d}V/\mathrm{d}T$ of the thermometer can be adjusted to be optimal for a certain temperature range by adjusting the %%@
bias current: With low bias--currents ($\sim$ 10 pA) there is a higher sensitivity at low--temperatures, while with higher %%@
bias--currents ($\sim$ 100 pA) sensitivity is gained at high temperatures, but lost at low temperatures. The best results with %%@
thermometry are obtained by repeating experiments with a few different bias points for different temperature ranges. However, %%@
higher bias currents may cause significant heating at the lowest operating temperatures due to power dissipation to the normal %%@
metal. The practical upper limit for the bias current $I$ depends on the resistivity of the normal metal material, but is at most %%@
$I$ corresponding to voltage $V_{\mathrm{SINIS}}\sim 2 \Delta/e$, after which significant heating is induced by the junction %%@
itself\cite{Jukkareview}. Overheating is a critical issue at low temperatures, especially with structures containing small normal %%@
metal islands or otherwise very well thermally isolated samples, such as suspended nanowires\cite{LT25,KoppinenPRL}. 
In addition to heating, the bias current can provide self--cooling of the normal metal island\cite{martiniscooler,leivocooler}. %%@
Usually, SINIS junctions designed for thermometry have small junction areas and therefore small self--cooling effects, that can be %%@
neglected in the analysis (Fig. \ref{ExptIV} (a)). However, even for larger junctions, the self--cooling can be avoided by %%@
choosing a proper bias point that never yields voltages near the optimal cooling point $V_{\mathrm{SINIS}}\sim 2\Delta/e$. 

Notice also from Fig. \ref{ExptIV} (b) that typically the measured SINIS response as a function of the refrigerator (bath) %%@
temperature deviates from the theory calculation (Eq. (\ref{current_sym}) below) at $T<150$ mK. This deviation is most likely 
due to noise heating of the electron gas, i.e. coupling of the electron gas into its electromagnetic environment causing overheating %%@
of the electrons so that $T_{\mathrm{electron}}\neq T_{\mathrm{bath}}$. This noise heating power, typically $\sim$ 10 fW is dependent on the filtering %%@
of the lines in the cryostat and the electrical impedance of the junctions.   

The theoretical single particle current of a NIS junction derived from the tunneling Hamiltonian is given by\cite{Tinkham}
\begin{equation}\label{BCS}
I(V,T)=\frac{1}{e R_{T}}\int_{-\infty}^{\infty}n_{S}(E)\left[f_{N}(E-eV,T)-f_{S}(E,T)\right]dE
\end{equation}
where $R_{T}$ is the tunneling resistance of the sample, $n_{S}(E)$ the density of states (DOS) of the superconductor and %%@
$f_{N,S}$ the Fermi--Dirac distributions in the normal metal and the superconductor, respectively.
Interestingly, Eq. (\ref{BCS}) can be rewritten in a symmetric form
\begin{equation}\label{current_sym}
	I(V,T)=\frac{1}{2e R_{T}}\int_{-\infty}^{\infty}n_{S}(E)\left[f_{N}(E-eV,T)-f_{N}(E+eV,T)\right]dE
\end{equation}
where the Fermi--distribution of the superconductor is eliminated. 
Hence, the $I$--$V$ is dependent only on the electron temperature of the normal metal, i.e. the NIS--junction probes directly the electron %%@
temperature of the normal metal and no additional knowledge of the temperature of the superconductor is needed. 
However, there is still an implicit dependence of the temperature of the superconductor, since the DOS depends on the superconducting %%@
gap, whose temperature dependence at $T>0.5T_{C}$ must be taken into account. Note also that if we relax the %%@
assumption of quasiequilibrium (Fermi--function distributions) for either terminal, this simplification is not valid anymore.

\begin{figure*}[ht]
\includegraphics[width=\textwidth]{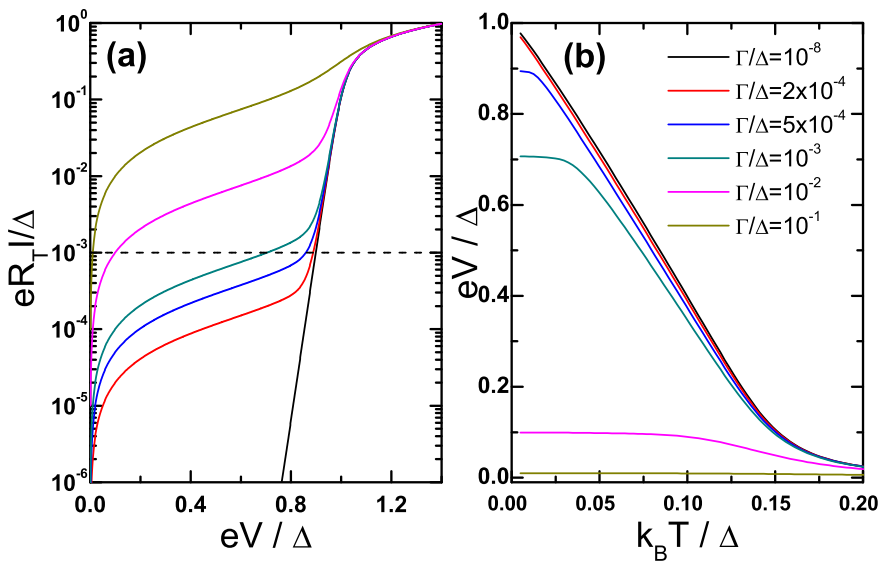}
\caption{\label{DynesFig}(Color online)(a) $I$--$V$ characteristics calculated at a temperature 50 mK with different values of the %%@
$\Gamma$ parameter (smallest to largest value of $\Gamma/\Delta$ corresponds to curves from bottom to top). Values are shown in %%@
Fig. (b). Horizontal, dashed black line corresponds to the current bias that has been used for calculating the calibration curves %%@
in (b). This correponds to 10 pA for junction with $R_{T}=$ 20 k$\mathrm{\Omega}$ and $\Delta=$ 220 \textmu eV. 
(b) SINIS thermometer voltage as a function of temperature, corresponding to the same values of $\Gamma$ as in (a) %%@
($\Gamma/\Delta$ increases from top to bottom).}
\end{figure*}

Typically, real junctions show a non--exponential finite sub--gap current (cf. Fig. \ref{ExptIV} (a) $I\sim$ 1 pA) at low enough %%@
temperatures and voltages. 
This sub--gap current can also be modelled with Eqs. (\ref{BCS})--(\ref{current_sym}) by incorporating a broadened DOS, i.e. by %%@
taking into account the finite life--time of quasiparticles in the superconductor \cite{Dynes}. The broadened DOS is then written %%@
as
\begin{equation}\label{DOS}
n_{S}(E)=\left|\mathrm{Re} \left\{ \frac{E+i\Gamma}{\sqrt{(E+i\Gamma)^2-\Delta^2}} \right\}\right| 
\end{equation}
where $\Gamma$ describes the magnitude of broadening of the DOS. Typically for evaporated Al films we have $\Gamma/\Delta \sim 2 %%@
\cdot10^{-4}$, consistent with values reported by other groups\cite{flyktman,oneil}. $\Gamma/\Delta$ is known to depend strongly on the %%@
material quality and can be two orders of magnitude larger for evaporated Nb films\cite{DynesNb}. 

Figures \ref{DynesFig}(a) and (b) show a calculation based on Eq. (\ref{current_sym}), how the broadening of the DOS affects the %%@
$I$--$V$ characteristics and voltage--to--temperature response of a SINIS. It is clear that the broadening of the DOS starts to %%@
influence the SINIS response mostly at low temperatures when $\Gamma/\Delta\sim 5\cdot 10^{-4}$, depending, of course, on the bias %%@
current value. At high biases, the broadening can be neglected to rather high values of $\Gamma$ but at lower biases the effect is %%@
stronger. The broadened DOS shows up in the $V_{\mathrm{SINIS}}$ vs. $T$ measurement qualitatively in the same way as noise heating, i.e. %%@
by leading to a saturation of the voltage at low temperatures. However, the two effects can be typically resolved in the %%@
experiment because the value for $\Gamma$ can be determined from the $I$--$V$ characteristics, i.e. $\Gamma$ changes the current deep in the sub--gap, whereas $T$ changes the slope of the current rise at the gap edge (cf. Fig. \ref{DynesFig}(a)). Typically this broadening is not a problem for thermometry in %%@
evaporated Al films, because the bias point can be taken above the sub--gap current. However, if the SINIS thermometers are %%@
fabricated e.g. from evaporated Nb films where $\Gamma/\Delta$ may be relatively high, broadening can be a real problem  reducing %%@
the responsivity of the thermometers.
%\label{sec:1}
%and \cite{RefJ}
%\subsection{Subsection title}
%\label{sec:2}
%as required. Don't forget to give each section
%and subsection a unique label (see Sect.~\ref{sec:1}).
%\paragraph{Paragraph headings} Use paragraph headings as needed.
%\begin{equation}
%\end{equation}

\section{Tunneling current with charging energy}
Charging effects (Coulomb blockade) due to small capacitances of the tunnel junction have an effect on thermometry, especially in %%@
the limit $E_{C}>>k_{B}T$. We consider the case of two identical junctions in series (with the same capacitances $C$ and tunneling %%@
resistances $R_{T}$), i. e. a symmetric SINIS thermometer. In addition, we discuss the case where a gate electrode is in close %%@
proximity with the normal metal island, so that the geometry is basically a  hybrid single electron transistor (SET) %%@
\cite{SINISSET1}. The $I$--$V$ characteristics of SINIS structures are naturally modified when the charging effects are taken into %%@
account, and can be controlled by applying a voltage $V_{g}$ to the gate\cite{singlecharge}. The tunneling rates through a single %%@
junction  with charging energy and biased with a voltage $V_i$ can be written as\cite{SINISSET2}
\begin{eqnarray}\label{CBtunnel}
	\Gamma^{i,+}(V,T,n)=\frac{1}{e^2R_{T}}\int_{-\infty}^{\infty}n_{S}(E)f_{S}(E,T_{S})[1-f_{N}(E-E_{n}^{i,+},T_{N})]dE \\
	\Gamma^{i,-}(V,T,n)=\frac{1}{e^2R_{T}}\int_{-\infty}^{\infty}n_{S}(E)f_{N}(E+E_{n}^{i,-},T_{N})[1-f_{S}(E,T_{S})]dE.
\end{eqnarray}
%when the charging energy is taken into account. 
Here the $E_{n}^{i,\pm}=\pm2E_{C}(n+n_{g}\pm0.5)\pm eV_{i}$ is the change in the electrostatic energy when an electron tunnels on %%@
to the island (+) and off the island (-) through junction $i=\{L,R\}$, where $L,R$ stand for left and right junction, %%@
respectively. $E_{C}\equiv e^2/(2C_{\Sigma})$is the charging energy, where $C_{\Sigma} \approx 2C+C_{g}$ is the total capacitance %%@
of the island with $C$ the junction capacitance and $C_{g}$ the gate capacitance, $en$ is the excess quantized charge ($n$ %%@
integer) on the island and $en_{g}=Q_{g}$ the offset charge, which can be be varied continuously by the gate electrode voltage. 

The current through the device can be calculated using these tunneling rates by solving a Master equation with the detailed balance condition 
\begin{equation}\label{DB}
	\Gamma^{+}(n)P(n)=\Gamma^{-}(n+1)P(n+1)
\end{equation}
where $\Gamma^{\pm}=\Gamma^{L,\pm}+\Gamma^{R,\pm}$ and $P(n)$ is the occupation probability of the corresponding charge state $n$ %%@
obeying the normalization condition $\sum_{n=-\infty}^{\infty}P(n)=1$.
Once this equation is solved, the current through the island can be calculated from the expression
\begin{equation} \label{Ec_current}
I_{i}(V,T)=-e\sum_{n=-\infty}^{\infty}P(n)[\Gamma^{+,i}(n)-\Gamma^{-,i}(n)]
\end{equation}
and $I=I_{L}=I_{R}$. It can be easily shown that Eq. (\ref{Ec_current}) reduces to Eq. (\ref{current_sym}) in the limit $E_{C}=0$. %%@
An interesting observation about the SINIS current with a charging energy included is that it cannot be written in a form that is %%@
fully independent of the superconductor temperature $T_{S}$, not even in the weak Coulomb blockade limit $E_{C}<k_BT$. Hence, the %%@
conclusion of Eq. (\ref{current_sym})  where a NIS junction can be used as a normal metal electron thermometer without additional %%@
knowledge of $T_S$ is only a special case when $E_{C} << k_BT$. This effect may be important when considering high charging energy %%@
devices used for cooling \cite{heattransistor}, where temperature differences between the normal metal and the superconductor 
occur because of the heat flow from the island to the leads. 
However, for typical temperature differences $\Delta T\sim 200$ mK achieved for Al based SINIS coolers,
the effect on the $I$--$V$ characteristics is only visible in the sub--gap region, and with typical bias currents it can be neglected. 

\section{Charging effects in thermometry}
\begin{figure*}[ht!]
\includegraphics[width=\textwidth]{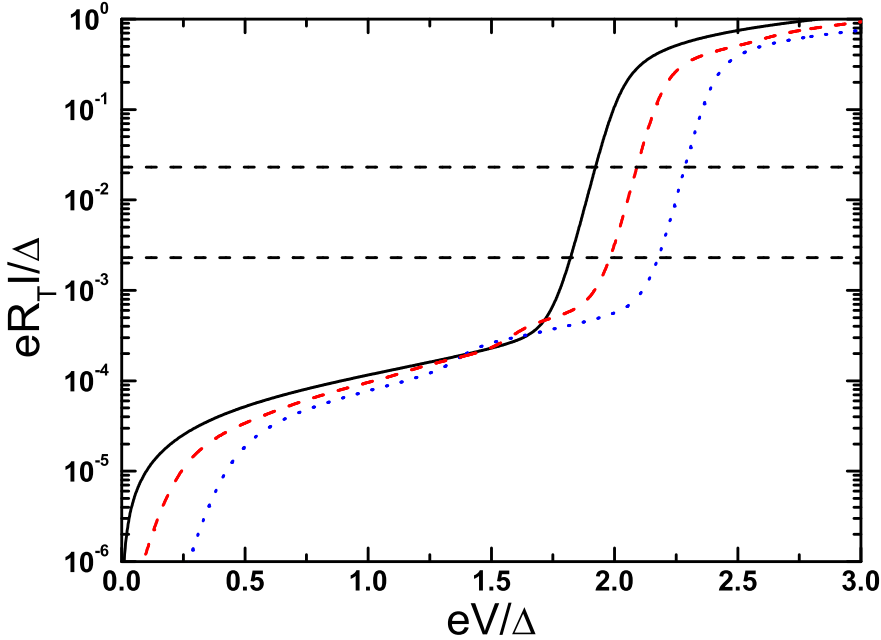}
\caption{\label{ECcurrent}(Color online) Calculated $I$--$V$ characteristics of a SINIS thermometer with three different values of %%@
$E_{C}/\Delta=0$ (black, solid), 0.1 (red, dashed) and 0.2 (blue, dot) and $n_{g}=0$. Two dashed horizontal lines correspond to the current bias %%@
values used  in later calculations 
$eR_{T}I/\Delta=2.3\cdot10^{-3}$ and $eR_{T}I/\Delta=2.3\cdot10^{-2}$.}
\end{figure*}

\begin{figure*}[ht!]
\includegraphics[width=\textwidth]{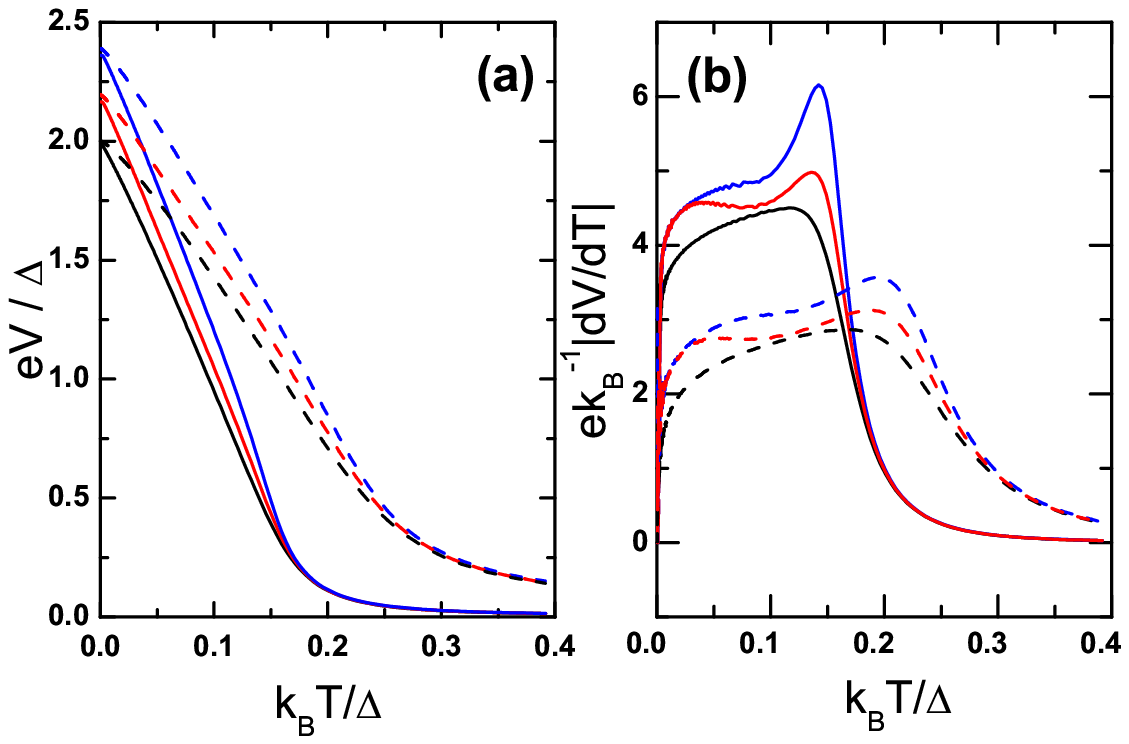}
\caption{\label{ECcalib}(Color online) (a) Calculated voltage response vs. temperature curves of a SINIS device for three %%@
different charging energies $E_{C}/\Delta=0$  (black), 0.1 (red) and 0.2 (blue), with $n_g=0$ (maximal Coulomb blockade). Solid lines %%@
are with low bias ($eR_{T}I/\Delta=2.3\cdot10^{-3}$) and dashed with high bias current ($eR_{T}I/\Delta=2.3\cdot10^{-2}$). (b) %%@
Responsivity $\left|\mathrm{d}V/\mathrm{d}T\right|$ calculated from (a).}
\end{figure*}

%Charging energy has an effect in usage of the SINIS structures as thermometers. 
%Calculated calibration curves with varying charging energy according to Eqs. (\ref{CBtunnel})--(\ref{Ec_current}) are in the %%@
%Fig. \ref{ECcalib}(a)
In this section we present the main theoretical results of the charging effects on thermometry and discuss how the gate can %%@
modulate the responsivity of the thermometer. All results are presented in scaled units and in all calculations the DOS broadening %%@
parameter $\Gamma/\Delta=2\cdot10^{-4}$ has been used. Sub--gap $I$--$V$ characteristics of a SINIS at temperature %%@
$k_{B}T/\Delta=0.02$ with three different values of charging energy $E_C/\Delta=$0, 0.1 and 0.2 are shown in Fig. \ref{ECcurrent}. %%@
It is clear that the charging energy effectively shifts the $I$--$V$ curves to higher voltages, leading to a behavior that %%@
resembles an effective increase of the superconducting gap from $\Delta$ to maximally $\Delta+E_C$ (if $n_g=0$). Furthermore, since the DOS is non--zero within the gap, a weak fingerprint of the Coulomb staircase can be seen in the $I$--$V$ curves in the sub--gap region at $eV/\Delta \sim 1.5$. 

In order to study how thermometry is affected, we should investigate how the voltage-to-temperature response is modified. This is %%@
shown in figure \ref{ECcalib}(a),  calculated from Eqs. (\ref{CBtunnel})--(\ref{Ec_current}) the same values of charging energy as %%@
used in calculating $I$--$V$ characteristics for Fig. \ref{ECcurrent}. The solid and dashed lines correspond to two different %%@
current bias values shown in Fig. \ref{ECcurrent}, where the low bias $eR_{T}I/\Delta=2.3\cdot10^{-3}$, and the high bias %%@
$eR_{T}I/\Delta=2.3\cdot10^{-2}$. These scaled current values correspond to 10 pA and 100 pA currents for junctions with %%@
$R_{T}=$ 50 k$\mathrm{\Omega}$ and $\Delta$ = 220 \textmu eV, typical for thin film Al. With this value of $\Delta$, the corresponding charging %%@
energies in Figs \ref{ECcurrent} and \ref{ECcalib} are $E_{C}/k_B=$0, 0.25 K and 0.5 K. The responsivities $\left|\mathrm{d}V/\mathrm{d}T\right|$ of %%@
the SINIS thermometer calculated from the voltage vs. temperature curves are presented  in Fig. \ref{ECcalib}(b). 

\begin{figure*}[ht!]
\includegraphics[width=\textwidth]{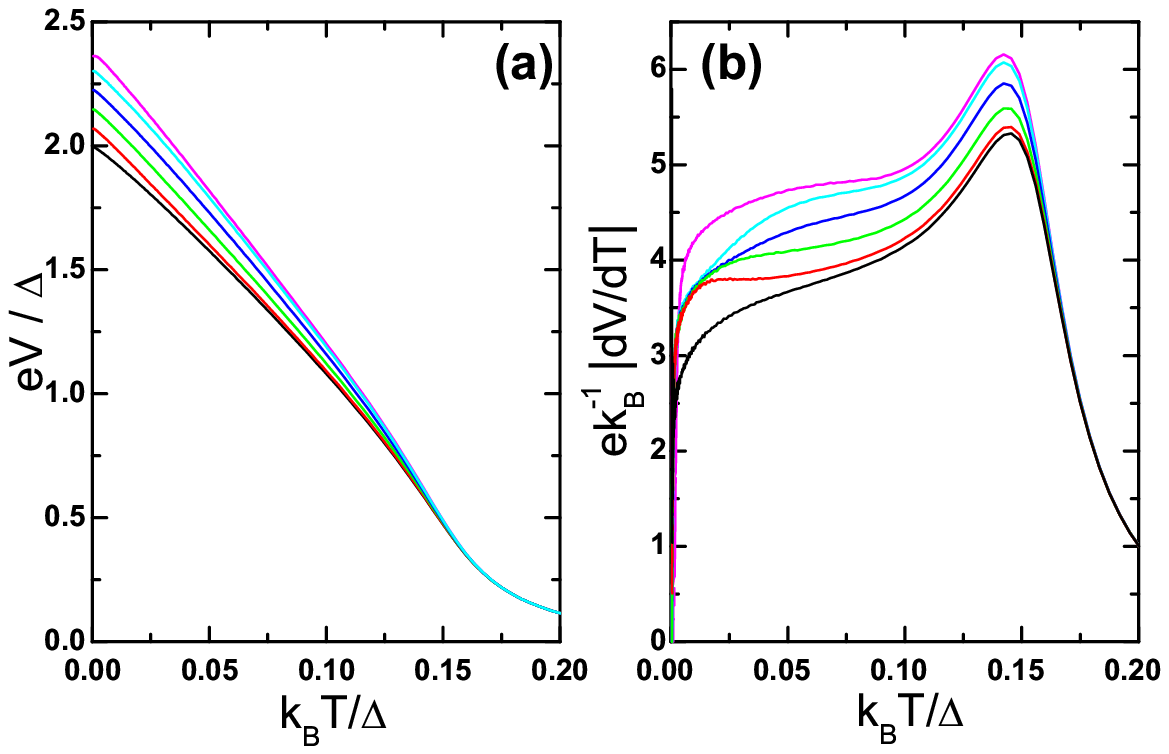}
\caption{\label{gatecalib}(Color online) (a) Calculated SINIS voltage vs. temperature curves for $E_C/\Delta=$ 0.2 and current bias %%@
$eR_{T}I/\Delta=2.3\cdot10^{-3}$ with different gate charge values varying from an open gate $n_g=0.5$(lowest, black curve)to a %%@
closed gate $n_g=0$ (magenta, top curve) with interval of $n_g=0.1$.  (b) Responsivities $\left|\mathrm{d}V/\mathrm{d}T\right|$ calculated from (a) %%@
corresponding to the same gate values (lowest black curve open gate and top magenta curve closed gate).}
\end{figure*}

It can be seen from Fig. \ref{ECcalib}(b) that the charging energy increases the responsivity of the thermometer, but it also %%@
changes the shape of the voltage vs. temperature curve: there appears a clear bump in the responsivity curve at around $k_BT=0.15 \Delta$ for the low bias (0.2$\Delta$ for the high bias) when the charging energy becomes appreciable. The shift of the 
effective gap is also clearly seen in Fig. \ref{ECcalib} (a) in the zero temperature limit, which moves from  $eV=2\Delta$ to %%@
$eV=2(\Delta+E_{C})$.
The responsivity curves also clearly show how the two different bias points have different optimal temperature ranges: at lower %%@
bias the thermometer has more responsivity for temperatures below the peak caused by charging effects ($k_B T/\Delta<0.15$, which %%@
corresponds to $T=380$ mK for the values of $\Delta$ and $R_T$ used above),  but quickly drops to unpractically small values above %%@
that.  However, at the higher bias the responsivity stays large up to much higher temperatures. 

\begin{figure*}[ht!]
\includegraphics[width=\textwidth]{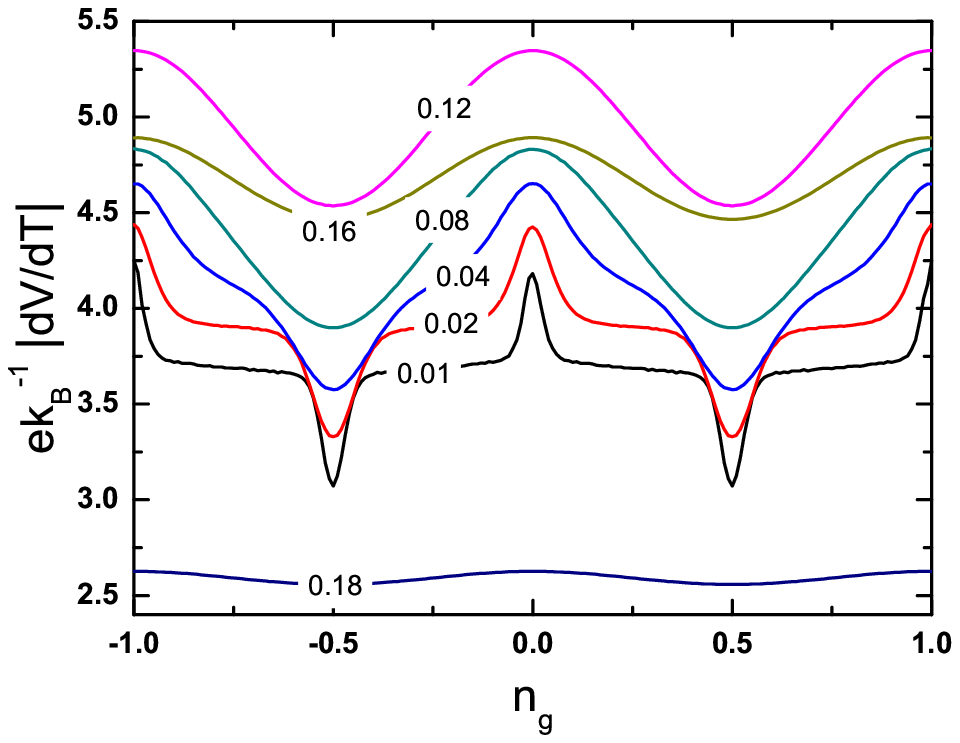}
\caption{\label{gatesweep}(Color online)Responsivity of a SINIS thermometer as a function of gate charge with different scaled %%@
temperatures from $k_{B}T/\Delta$= 0.01 to 0.18. The same charging energy and current bias values are used as in Fig. %%@
\ref{gatecalib}}
\end{figure*}

Gate modulation of the voltage response vs. temperature curves and corresponding responsivities are presented in Fig. %%@
\ref{gatecalib}(a) and (b). Here the highest value of $E_{C}/\Delta=$ 0.2 from the plot in Fig. \ref{ECcalib} is used, and all %%@
curves are calculated with the low current bias value $eR_{T}I/\Delta=2.3\cdot10^{-3}$. Open and closed gate situations correspond %%@
to the top (magenta) and bottom curves (black), respectively, and gate voltages between these two extrema are plotted  with an %%@
interval of $n_{g}=0.1$. The open gate ($n_g=0.5$), interestingly, has the same zero temperature limit $eV=2\Delta$ as a thermometer %%@
with zero charging energy, however the full temperature dependent responsivity is different. This can be seen most easily from the %%@
responsivity curves in Fig. \ref{gatecalib}(b), where we see that the responsivity bump caused by $E_C$ still exist. Hence, the %%@
shape of the $V$ vs. $T$ curve is still different.
Even more surprisingly, at low temperatures $k_BT < 0.02 \Delta$ the responsivities with intermediate gate voltage values seem to %%@
merge, while the open and closed gate values are still well separated.  This effect is more clearly shown in Fig.\ref{gatesweep}.

Figure \ref{gatesweep} shows the responsivity of the thermometer as a function of the gate charge number $n_{g}$ at different %%@
temperatures from $k_{B}T/\Delta$= 0.01 to 0.18. These curves were calculated with the same parameters as in Fig. \ref{gatecalib}. %%@
The low temperature results show sharply peaked values at the gate open and closed positions, while the intermediate values show %%@
flat regions, where the responsivity does not depend on the gate charge, as noted before. At higher temperatures, these sharp %%@
peaks broaden into a sinusoidal dependence whose amplitude decreases with increasing temperature. The largest change in the %%@
responsivity is obtained at the lowest temperatures, where responsivity can be enchanced by 30 $\%$ by closing the gate at %%@
$k_BT=0.01 \Delta$, for example.

\section{Relevance to experiments}
\begin{figure*}[ht!]
\includegraphics[width=\textwidth]{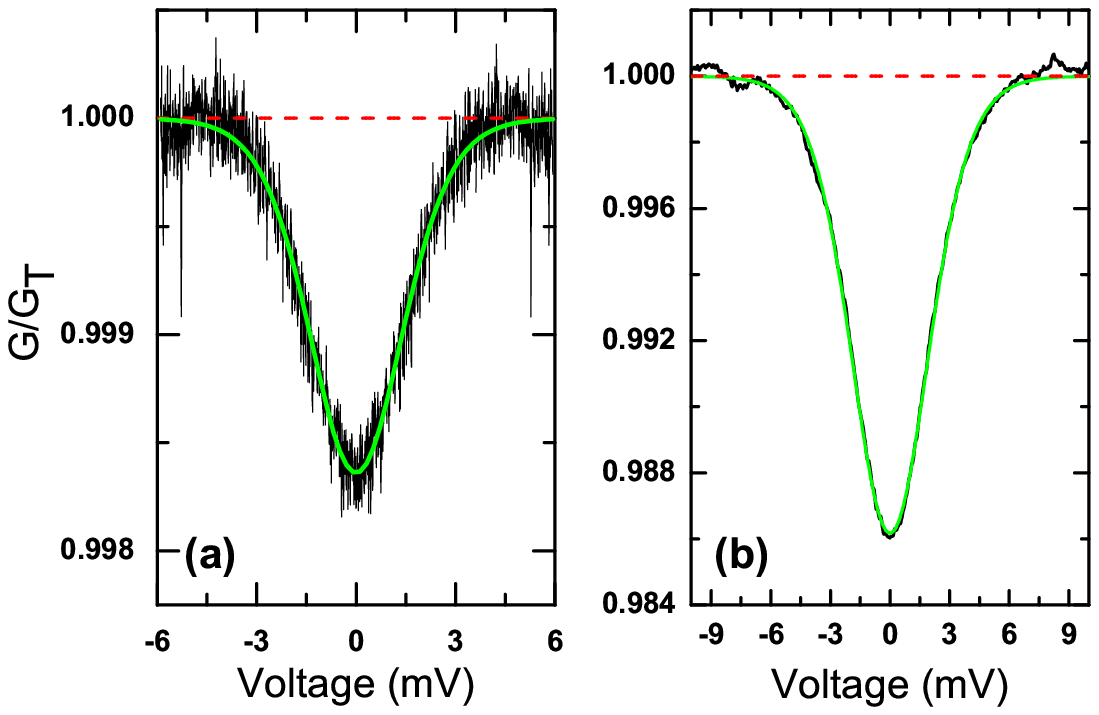}
\caption{\label{CBmeas}(Color online)The measured (black line) differential conductance spectrum of a large junction area SINIS %%@
device (0.35 \textmu m$^2$) (a) and a smaller device (junction area 0.05 \textmu m$^2$) (b) both at 4.2 K. Conductance is normalized with %%@
$G_{T}=1/R_T$ (red, dashed line). Solid (green/gray) line corresponds to a weak--Coulomb blockade regime fit to the data in the %%@
lowest order in $E_C/(k_B T)$ \cite{Jukkareview}.}
\end{figure*}

To decide in practice when one should consider Coulomb charging in SINIS thermometry, one needs to measure the value of the charging %%@
energy $E_C$ and compare it with the temperature range of interest. The $E_C$ measurement can easily be performed at 4.2 K in the  %%@
weak Coulomb blockade ($E_{C}<k_{B}T$) limit, where the size of the zero bias  Coulomb blockade dip  in the tunneling conductance %%@
spectrum $\Delta G$ depends on the charging energy by the relation \cite{Jukkareview}
\begin{equation}
\frac{\Delta G}{G_{T}}=\frac{E_{C}}{3k_{B}T}, 
\end{equation}
 where $G_{T}$ is the tunneling conductance around $V=0$ without the dip.  As an example,  
 figure \ref{CBmeas} shows two measured Coulomb blockade dips at 4.2 K with (a) a typical larger junction area (0.35 \textmu m$^2$) %%@
Al/Cu/Al SINIS device designed to act as a cooler \cite{LT25,KoppinenPRL}, and (b) a smaller typical solitary Al/Cu/Al SINIS thermometer with %%@
a junction area $\sim 0.05~$\textmu m$^2$.  From the measurements we obtain $E_{C}/k_B=$ 20 mK for the cooler sample, showing that it is %%@
in the limit of weak Coulomb blockade $k_{B}T>E_{C}$ for our experimentally achievable (dilution refrigerator) temperature range. %%@
The changes to the voltage-to-temperature response are within the experimental error in that case, and analysis can be carried out %%@
with the simpler theory of Eq. (\ref{current_sym}), without charging effects.

%The measured charging energy at 4.2 K for our cooler junction is shown in Fig. \ref{CBmeas}. 

%We have also calculated the BCS--theory curves with and without charging energy (not shown here) and condirmed, that it has no %%@
%effect on calibration. 

However, the effects of charging energy are more observable in the second, smaller SINIS thermometer (5 \textmu m x 300 nm x 30 nm %%@
normal metal island), with a measured charging energy  $E_{C}/k_B\sim$ 200 mK (Fig. \ref{CBmeas} (b)).  This type of thermometer is %%@
typically used e.g. for probing the bath or (local)phonon temperature during an experiment. The measured $E_C$ corresponds to a %%@
value $E_C/\Delta \sim 0.08$ for Al, which leads to observable changes in the temperature response based on the calculations in %%@
the previous section.  

\section{Conclusions}
We have studied the charging effects on SINIS tunnel junction thermometry and shown that for small enough junctions, the %%@
responsivity of the thermometer can be modulated with an additional gate electrode in close proximity to the junctions, with maximum %%@
responsivity achieved with the closed gate situation. In addition, the shape of the voltage response vs. temperature curve changes %%@
when the charging energy is taken into account,  leading to the conclusion that in typical solitary SINIS thermometers  charging %%@
effects must be taken into account in the conversion from measured voltage to temperature (calibration), and have to be computed %%@
numerically. The effect of an unknown offset charge will not lead to ambiguities in the analysis, as the shape of the voltage vs. %%@
temperature curve changes as a function of the offset charge, so that the value of the offset charge can be determined %%@
self-consistently.

%\begin{figure}
%  \includegraphics{example.eps}
%\end{figure}
%\begin{figure*}
%  \includegraphics[width=0.75\textwidth]{example.eps}
%\end{figure*}
%\begin{table}
%\begin{tabular}{lll}
%\hline\noalign{\smallskip}
%\end{tabular}
%\end{table}

%\acknowledgement

\begin{acknowledgements}
P.J.K. acknowledges Magnus Ehrnrooth foundation and T.K. acknowledges Emil Aaltonen foundation for financial support. We 
thank J. Pekola for discussions. This work has been supported by the Academy of Finland projects No. 118665 and 118231.
\end{acknowledgements}

%\bibliographystyle{apsrev}
%\bibliography{CBSINIS}
%\bibliographystyle{plain}

%\begin{thebibliography}{}
%\bibliographystyle{natbib}
%\bibliographystyle{plain}
% etc
%\end{thebibliography}

\end{document}